\documentclass{svjour3-a} 
\usepackage{amssymb}
\usepackage{amsmath}
\usepackage{graphicx}
\usepackage[authoryear]{natbib}
\journalname{Arxiv preprint}
\usepackage[a4paper,inner=1in,outer=1in,top=1.3in,bottom=1.3in]{geometry}

\input{tcilatex}
\begin{document}

\def\inputp[#1]#2{\def\inputpar{#1}\input{#2}}
\def\citel#1{\citeauthor{#1}, \citeyear{#1}}

\newcommand{\negpar}[1][-1em]{%
  \ifvmode\else\par\fi
  {\parindent=#1\leavevmode}\ignorespaces
}


\title{Graph clustering in industrial networks}
\author{V. Bouet \and A.Y.Klimenko}
\date{March 2019}
\institute{SoMME, The university of Queensland, Qld 4072, Australia, email: klimenko@mech.uq.edu.au}

\maketitle

\begin{abstract}
{The present work investigates clustering of a graph-based representation of industrial connections derived from international trade data by Hidalgo et al (2007) and confirms existence of around ten industrial clusters that are reasonably consistent with expected historical patterns of diffusion of innovation and technology. This supports the notion that technological development occurs in sequential innovation waves. The clustering method developed in this work follows conceptual ideas of Lambiotte and Barahona (2009), who suggested to use random walk to assess a hierarchical structure of network communities where different levels of the hierarchy correspond to different diffusion times. We, however, implement these ideas differently to match physics of the problem under consideration and introduce a hierarchal clustering procedure that is combined with convenient resorting of the elements. An equivalent spectral interpretation of the clustering is also given and discussed in the paper. }  
{Graph clustering, random walk, diffusion of innovation and technology}   
\end{abstract}


\section{Introduction}

The classical view of economics, which can be traced back to Adam Smith and
David Ricardo, is firmly based on equilibrating economic forces presuming
that, when disturbed, economic equilibrium is promptly restored by these
forces \citep{Freeman1997}. There is, however, an alternative view
introduced by \citet{Schumpeter1947}, who explained the existence of
Kondratiev cycles in economic activity by a sequence of technological
revolutions. These revolutions are instigated by a surge of innovations that
move economy away from equilibrium by eliminating obsolete technologies and
thus creating conditions favorable for rapid economic progress (i.e.
Schumpeter's concept of "creative destruction"). After a technological
surge, higher profit margins tend to persist in innovative sectors for a
substantial time until these sectors reach maturity and the wave fades away.

While, according to the classical interpretation, economy drives
technological development, the alternative view is that technological
progress plays the leading role over longer periods of time and create
technological waves modulating economic growth. The inquisitive reader may
note that technological waves of appreciable magnitude are possible only if
different innovations are somehow connected to each other --- a single
innovation would rapidly blend into the economy and cannot create a lasting
disturbance of the equilibrium. Existence of industrial clusters is, thus, a
principal question that may pose a strong argument in favour of the
Schumpeterian interpretations. \citet{Atlas2007} and \citet{Atlas2014} have
recently introduced a theory and a method that allows us to examine links
and connections between different industries by analysing global export
data. This theory relates industrial connections to international trade data
and demonstrates that industrial development tends to take place within
connected proximities of existing industries \citep{Atlas2007}. This process
can be interpreted as diffusion of resources and innovation between
connected sectors of industry, which can be modelled by a random walk on
graphs representing industrial connections. The fact that these data are
readily available from the United Nations databases makes a strong argument
in favour of this method. The links between industries seem to indicate the
presence of industrial clusters (see the graph shown in the next chapter).
The goal of the present work is a more rigorous and formal analysis of the
existence and structure of clusters for the undirected industrial network
introduced by \citet{Atlas2007} and \citet{Atlas2014}.

A large number of publications is dedicated to graph partitioning and
identification of communities %
\citep[][]{clust2009,Mucha2009,clust2010,clust2015,clust2016,clust2017}. The
spectral method, based on using the Fiedler vector, seems to be the most
popular but, as \citet{clust2016} remark, the problem of identifying network
communities is ill-defined --- there is no universal definition or algorithm
that determines communities in some unique or undisputedly superior way. The
method of identifying communities needs to reflect a physical understanding
of the problem. In the present case, this implies the involvement of random
walks, since diffusion on the network is expected to reflect the process of
diffusion of innovation within and between the industrial clusters %
\citep[][]{Atlas2007}. The literature dedicated to random walks is vast; %
\citet{graphRW2014} and \citet{graphRW2017} presented the excellent reviews
of random walks on graphs, while \citet{graphProb2018} overviews a broader
spectrum of issues associated with probabilistic processes in networks. In
general, it would be productive to consider time-dependent or multisliced
networks %
\citep[][]{clust2010,HolmePetter2012Tn,Mucha2013,t_graph2015,Tnet2016},
since links between industries obviously evolve in time. However, detailed
historical information about the evolution of these links does not exist
and, therefore, all links are necessarily treated as time-independent.

In many respects, the present analysis is similar to the concept of
partition stability introduced in \citet{Lamb2009} and\ extended in more
recent publications \citet{PNAS2010,Schaub2012,Lamb2015} to involve both
discrete and continuous time as well as other generalisations %
\citep[e.g.][]{NatureC2014,srep2016}. As suggested by \citet{Lamb2015}, we
use discrete-time random walks (i.e. a Markov chain) to analyse the
community structure of a network, which corresponds to different resolution
levels for different diffusion times. This approach allows us to examine
hierarchies of communities, which have been discussed in a few publications
that, generally, may or may not be related to random walks %
\citep[e.g.][]{clust2009b,clust2010,clust2015,Lamb2015,clust2017}. There
are, however, some essential differences between our work and the partition
stability approach. First, we understand and define clusters differently
from \citet{Lamb2009} --- this is determined by our physical interpretation
of the industrial networks introduced by \citet{Atlas2007}. This difference
is explained further in Section \ref{SecRW}. Second, the stability of a
partition is suggested by \citet{Lamb2009} as a criterion, which is designed
to assess the quality of a community partition produced by other algorithms,
while we use our approach as both a definition of clusters and an algorithm
producing a hierarchy of partitions and a convenient ordering of the nodes.

Our approach has both useful transitional and conceptually transparent
spectral interpretations. It allows us to identify nine primary industrial
clusters, which are reported in the last sections of this work. Although
designed for a specific problem, the clustering procedure introduced in the
present work is generic and, at least in principle, can be used in different
applications (and, possibly, with different clustering criteria). This
procedure is not intended to enforce the fastest possible division of a
graph into a given number of clusters, but to follow the mechanics of
diffusion of innovation explained above, and examine the complex structure
of connections created by this diffusion.

\section{Proximity of industries and international trade.}

The theory of proximity of different industrial products, introduced by %
\citet{Atlas2007}, statistically reflects correlations between success of a
country in trading a certain product and success in trading other products
that are technologically associated with the first product. If $X_{\alpha j}$
represents export of product $j$ by country $\alpha $ then the quantity 
\begin{equation}
\tilde{X}_{\alpha j}=\frac{1}{\tilde{Y}_{\alpha }}\frac{X_{\alpha j}}{%
\sum_{\beta }X_{\beta j}},\ \ \ \tilde{Y}_{\alpha }=\frac{Y_{\alpha }}{%
\sum_{\beta }Y_{\beta }},\ \ Y_{\alpha }=\sum_{j}X_{\alpha j}
\end{equation}%
which is called revealed competitive advantage, represents the fraction that
a particular country $\alpha $ contributes to the world trade in exporting a
particular product $j$ related to $\tilde{Y}_{\alpha }$ --- the fraction of
all exports of this country in the world trade. Values of $\tilde{X}_{\alpha
j}\geq 1$ indicate that country $\alpha $ has a comparative advantage in
exporting product $j$. The index-function 
\begin{equation}
M_{\alpha j}=\left\{ 
\begin{array}{c}
1,\ \ \tilde{X}_{\alpha j}\geq 1 \\ 
0,\ \ \tilde{X}_{\alpha j}<1%
\end{array}%
\right. 
\end{equation}%
indicates whether country $\alpha $ is competitive in exporting product $j$.
Proximity, which is defined by 
\begin{equation}
\varphi _{ji}=\sum_{\alpha }\frac{M_{\alpha j}M_{\alpha i}}{\max
(K_{j},K_{i})},\ \ \ K_{j}=\sum_{\alpha }M_{\alpha j}  \label{proxi}
\end{equation}%
is similar to the matrix of correlation coefficients between columns of the
matrix $M_{\alpha j}$. Proximity defined by (\ref{proxi})\ is symmetric and
constrained $0\leq \varphi _{ji}\leq 1$ where $\varphi _{ji}=1$ only if the
columns $j$ and $i$ are the same. Large values of the proximity $\varphi
_{ji}\sim 1$ indicate that a country producing product $j$ also tends to
produce product $i$ and, hence, products $i$ and $j$ are very likely to be
related. Small values of $\varphi _{ji}$ do not tell us anything specific
since small variations of $\varphi _{ji}$ are likely to be coincidental, not
revealing anything about real-world links between the products. We wish to
consider only connections between products that have largest proximities and
ignore proximities of smaller magnitudes --- only large values of $\varphi
_{ji}$ are indicative of a technological or logistic connection between the
products. This, however, would result in a disconnected graph. Therefore, we
follow \citet{Atlas2007} and use a two-stage procedure. First, Kruskal's
algorithm is used to connect $N$ product nodes by a tree with $N_{1}=N-1$
edges. This algorithm selects largest proximities one by one and connects
the corresponding nodes only if this connection does not form a loop. In the
second stage, the $N_{2}$ largest values $\varphi _{ji}$, that were not used
as connectors in the first stage, are selected to form additional
connections. The result is a complex undirected graph, which is shown in
Figure \ref{figa}. This figure seems to indicate presence of 6 clusters
outlined by red rectangles. The present work investigates the existence of
clusters in this graph by introducing and using a more rigorous procedure
for cluster analysis.

\section{Defining clusters by using a random walk \label{SecRW}}

Our definition of clusters is necessarily based on random walks on the graph
specified in the previous section since it is this random walk that,
according to \citet{Atlas2007}, emulates the rate of diffusion of
technological innovations between industrial sectors. Mathematically, random
walk is represented by a discrete Markov chain \citep[see][]{AMSprob1997}),
which is characterised by single-step transitional probabilities $P_{ji}.$
The distribution of probabilities at time step $t$ denoted by $p_{i}(t)$
satisfies 
\begin{equation}
p_{j}(t+1)=\sum_{i}P_{ji}p_{i}(t)\text{ or }\mathbf{p}(t+1)=\mathbf{P\,p}(t)
\end{equation}%
if the vector - matrix notation is used. The transitional probabilities are
specified by 
\begin{equation}
\mathbf{P=}(1-\beta )\mathbf{T+}\beta \mathbf{I,\ \ \ T=AD}^{-1}
\label{Pbet}
\end{equation}%
where $\mathbf{A}$ is an $N\times N$ symmetric adjacency matrix: $A_{ji}=1$
if nodes $i$ and $j$ connected and $A_{ji}=0$ otherwise ($A_{ii}=0$), $T$ is
its version subject to the probability-preserving normalisation, $\mathbf{D}$
is a diagonal matrix with $D_{ii}=d_{i}$ specifying the degree of node $i$, $%
\mathbf{I}$ is the unity matrix, and $N$ is the number of nodes in the
graph. The parameter $\beta $ specifies the probability of a randomly
walking particle to remain at the same node. As discussed in the next
section, $\beta =1/2$ is a suitable, safe choice for this parameter. The
normalisation of the transition matrix preserves the overall probability $%
\Sigma _{i}p_{i}=1$. It is possible to consider random-walk transition
probabilities proportional to the proximities $\varphi _{jk}$ of the
connected nodes but this would not substantially affect the results and we
use the simpler definition of the adjacency matrix given above. The matrix $%
P_{ji}^{n}$ defined by 
\begin{equation}
\mathbf{P}^{n}\mathbf{=}\underset{n\text{ times}}{\underbrace{\mathbf{%
P\,P\,...\,P}}}
\end{equation}%
specifies the probability distributions $\left( p_{j}\right)
_{i}^{n}=P_{ji}^{n}$ evolved from the initial localisation at node $i$ after 
$n$ time steps.

The definitions of partition stability given by \citet{Lamb2015} interpret
clusters as subgraphs that tend to retain a random walk for a given number
of time steps. This definition involves maximisation of stability --- a
certain quantity defined on the basis of the $n$-step transitional
probabilities $P_{ji}^{n}$ and stationary distributions $p_{j}^{\circ
}\equiv d_{j}/(2E),$ where $E$ is the number of edges. In simple terms, a
partition into clusters after $n$ time steps is considered to be of good
quality when, on average, $P_{ji}^{n}$ is relatively large (above the
stationary distribution when properly scaled) if $i$ and $j$ belong to the
same cluster, and is relatively small if $i$ and $j$ are not from the same
cluster. That implies that clusters are expected to have minimal
interactions. Our understanding of clusters does not require that members of
an industrial cluster must not significantly interact with other industries
(which would not seem justified) but that, after $n$ steps, these
interactions become similar for all members of the same cluster. Therefore,
after formation of a cluster, the cluster members are expected to display
similar evolutionary dynamics.

Our interpretation of clustering can be expressed in terms of the similarity
relation given in the following proposition

\begin{proposition}
Two nodes $i$ and $j$ are deemed to be similar $\mathbf{p}_{i}^{n}\,\cong 
\mathbf{p}_{j}^{n}$ and belong to the same cluster at the time step $n$
provided the probability distributions originated from these nodes (i.e. $%
\mathbf{p}_{i}^{n}\,$\ and $\mathbf{p}_{j}^{n}$) are similar for this and
the subsequent time steps.
\end{proposition}

The similarity, which is denoted by $\mathbf{p}_{i}^{n}\,\cong \mathbf{p}%
_{j}^{n}$ here and in the rest of the paper, can be defined in different
ways. The definition 
\begin{equation}
R_{ij}^{n}=\mathbf{q}_{i}^{n}\cdot \mathbf{q}_{j}^{n}=\left( \mathbf{q}%
_{i}^{n}\right) ^{\func{T}}\left( \mathbf{q}_{j}^{n}\right) \geq 1-\,\delta 
\text{, \ \ }\left( q_{k}\right) _{i}^{n}\equiv \text{\ }\sqrt{\left(
p_{k}\right) _{i}^{n}}  \label{S1}
\end{equation}%
with a sufficiently small threshold $\,\delta $ is used due to its practical
stability and simplicity of avoiding the need to renormalise vectors since $%
\mathbf{q}_{i}^{n}\cdot \mathbf{q}_{i}^{n}=1$ for any $i$ because of the
probability normalisation. Conceptually, this or any other choice evaluating
a quantity similar to a correlation coefficient between $\mathbf{p}%
_{i}^{n}\, $\ and $\mathbf{p}_{j}^{n}$ would be suitable to define
similarity $\mathbf{p}_{i}^{n}\,\cong \mathbf{p}_{j}^{n}$.

Note that the initial conditions at $t=0$ correspond to $\mathbf{P}^{0}=%
\mathbf{I}$ and, consequently, $R_{ij}^{0}=0.$ \ Assuming that the graph
under consideration is connected and $\mathbf{P}$ is irreducible, the
opposite limit of $t=n\rightarrow \infty $ implies that $\mathbf{P}%
^{n}\rightarrow \mathbf{P}^{\infty }=\left[ \mathbf{p}^{\circ },...,\mathbf{p%
}^{\circ }\right] .$ That is distributions become stationary irrespective of
the initial localisation: $\mathbf{p}_{i}^{n}\rightarrow \mathbf{p}^{\circ }$%
\ or, with the use of the component notations, $P_{ji}^{n}\rightarrow
p_{j}^{\circ }\equiv d_{j}/(2E)$ for any $i$ and $n\rightarrow \infty $.
Note that $R_{ij}^{n}\rightarrow 1$ as $t=n\rightarrow \infty $ since all
columns of $\mathbf{P}^{n}$ become fully correlated. In this work, we
consider only connected graphs since, otherwise, clustering can be performed
independently for each of the connected components. The graphs under
consideration are undirected (and thus satisfy the detailed balance $%
P_{ji}p_{i}^{\circ }=P_{ij}p_{j}^{\circ }$) but, in practice, the procedure
given here may tolerate some degree of directionality. The details of this
procedure are considered further in this section .

\bigskip

The original ordering of the nodes may not be the best to represent cluster
hierarchies. The clustering procedure is accompanied by reordering of the
elements. For the purposes of this procedure, single nodes are considered to
be clusters containing a single element. Clusters, say $\mathfrak{A}$ and $%
\mathfrak{B}$ that have the corresponding numbers of elements (nodes) $N_{%
\mathfrak{A}}$ and $N_{\mathfrak{B}},$ are considered to be similar $%
\mathfrak{A\cong B}$ provided 
\begin{equation}
\bar{R}_{\mathfrak{AB}}^{n}\equiv \frac{1}{N_{\mathfrak{A}}N_{\mathfrak{B}}}%
\tsum\limits_{i\in \mathfrak{A}}\tsum\limits_{j\in \mathfrak{B}%
}R_{ij}^{n}\geq 1-\delta  \label{SC}
\end{equation}%
where $R_{ij}^{n}$ is defined by (\ref{S1}). Jointly with (\ref{SC}), this
definition somewhat resembles the Frobenius inner matrix product. At $N_{%
\mathfrak{A}}=N_{\mathfrak{B}}=1,$ definition (\ref{SC}) is obviously
consistent with the previously defined similarity of the nodes. The limiting
value $\bar{R}_{\mathfrak{AB}}^{n}=1$ is achieved if and only if the
clusters $\mathfrak{A}$ and $\mathfrak{B}$ are composed of identical
elements $\mathbf{p}_{i}=\mathbf{p}_{j}$ for all $i,j\in \mathfrak{A}\cup 
\mathfrak{B}$. Note that merging preserves self-similarity of the clusters,
that is, if $\bar{R}_{\mathfrak{AB}}^{n}\geq 1-\delta ,$ $\bar{R}_{\mathfrak{%
AA}}^{n}\geq 1-\delta $ and $\bar{R}_{\mathfrak{BB}}^{n}\geq 1-\delta ,$
then $\bar{R}_{\mathfrak{CC}}^{n}\geq 1-\delta $ where $\mathfrak{C=A}\cup 
\mathfrak{B}$, \ $\mathfrak{A}\cap \mathfrak{B=\varnothing }$ and $N_{%
\mathfrak{C}}=N_{\mathfrak{A}}+N_{\mathfrak{B}}$. Indeed, 
\begin{eqnarray*}
1 &\geq &\bar{R}_{\mathfrak{CC}}^{n}=\frac{N_{\mathfrak{AA}}^{2}\bar{R}_{%
\mathfrak{AA}}^{n}+2N_{\mathfrak{A}}N_{\mathfrak{B}}\bar{R}_{\mathfrak{AB}%
}^{n}+N_{\mathfrak{BB}}^{2}\bar{R}_{\mathfrak{BB}}^{n}}{N_{\mathfrak{C}}^{2}}
\\
&\geq &\underset{=1}{\underbrace{\frac{N_{\mathfrak{AA}}^{2}+2N_{\mathfrak{A}%
}N_{\mathfrak{B}}+N_{\mathfrak{BB}}^{2}}{N_{\mathfrak{C}}^{2}}}}\underset{%
\geq 1-\delta }{\ \underbrace{\min \left( \bar{R}_{\mathfrak{AA}}^{n},\bar{R}%
_{\mathfrak{AB}}^{n},\bar{R}_{\mathfrak{BB}}^{n}\right) }}\geq 1-\delta
\end{eqnarray*}

The clusters that are similar are merged with preservation of the original
ordering within and between the merged clusters. The overall ordering of
elements, which is represented by ordering of the clusters and ordering of
elements within the clusters, is nevertheless changed as the clusters merge:
subsequent similar clusters are moved to their first similar cluster to
achieve a merger. Note that the similarity of the clusters is, generally,
not transitive, that is 
\begin{equation}
\mathfrak{A}\cong \mathfrak{B\cong C}\ncong \mathfrak{A}  \label{intrans}
\end{equation}%
is a possibility for some set of clusters $\mathfrak{A},$ $\mathfrak{B}$ and 
$\mathfrak{C}$. That is, $\mathfrak{A}$ and $\mathfrak{C}$ are both similar
to $\mathfrak{B}$ and must be reasonably similar to each other but may or
may not be similar according to formal definition (\ref{SC}). If $\mathfrak{C%
}\ncong \mathfrak{A}$, the outcome of merging of $\mathfrak{A},$ $\mathfrak{B%
}$ and $\mathfrak{C}$ is dependent on the initial ordering (i.e. whether $%
\mathfrak{A}$ and $\mathfrak{B}$ or $\mathfrak{B}$ and $\mathfrak{C}$ are
merged first). However, the algorithm we use forms clusters independently of
the initial ordering of the nodes. This can be achieved: 1) by transitive
closure of the similarity relation (i.e. enforcing $\mathfrak{C}\,\cong 
\mathfrak{A}$ in example (\ref{intrans})) or 2) by ordering similarities
between clusters, performing merges of more similar clusters first and then
re-evaluating similarity. Transitive closure is more simple and, by default,
is used in the present work. Hence, the practical algorithm used here allows
for slight compromises over condition (\ref{SC}) and this does not seem to
cause any practical problems (multiple mergers are infrequent and violations
of (\ref{SC}) are small).

The algorithm considered here does not produce a unique ordering of the
nodes --- the final ordering remains dependent on the initial ordering.
Clusters are formed independently of ordering of the elements and different
orderings produced by the algorithm are suitable to visualise clusters (best
appearance is a subjective matter). If the requirement of producing a unique
ordering at the end is imposed, the nodes need to be pre-ordered by another
fixed algorithm --- say by using ordering of the conventional Fiedler vector
--- and then further ordered by the clustering algorithm. The Fiedler vector
corresponds to second smallest eigenvalue (non-zero in this case) of the
Laplacian matrix $\mathbf{L=D-A}$ \citep{clust2009}.\ Note that the Fiedler
vector outlines the least connected (and therefore the slowest converging)
component of the graph while the clustering algorithm considers a hierarchy
of clusters that correspond to different characteristic times.

The procedure specified in this section has not been optimised for speed and
various measures can be implemented to speed up the simulations. For
example, replacing $\mathbf{P}^{n+1}\mathbf{=\mathbf{P}P}^{n}$ by $\mathbf{P}%
^{2n}\mathbf{=\mathbf{P}}^{n}\mathbf{P}^{n}$ would double the time step or,
generally, $R_{ij}^{n}$ does not need to be evaluated for all $i$ and $j.$
The speed, however, was not a pertaining issue in the present simulations.

The physical interpretation of clustering is transparent --- two nodes
belong to the same cluster if they are well-connected and the distinction
between random walks originated at these nodes disappears at a given time
step. This definition, obviously, depends on time: as clusters grow and
merge, they generally are different at different time steps (merges of
clusters introduce cluster hierarchies that are discussed further in the
paper). Therefore, clustering is not absolute but depends on characteristic
times of observation of the diffusive processes. The clustering algorithm,
which is constructed on the basis of the definition of node similarity given
above, primarily introduces a convenient reordering of nodes that makes
clustering structure of the graph visible. This reordering is fully
algorithmic. The cluster structure is then represented well by the
clustering map --- a plot of significant clusters versus diffusion time. As
clusters evolve in time, there is some freedom in selecting the resulting
compositions of the clusters --- we expect that a well-defined cluster
exceeds some minimal size and remains invariant (or approximately invariant)
over some range of characteristic times.

\section{Spectral representation of the clusters}

This section gives some additional explanations and introduces an
alternative interpretation of clustering based on spectral expansions. This
interpretation seems to be useful for a more rigorous conceptual
understanding of clustering, while the clustering algorithm of the previous
section seems to be more convenient as an engineering tool. The spectral
interpretation is based on the following lemma

\begin{lemma}
\label{lem1}The matrix $\mathbf{P}^{n}=[\mathbf{p}^n_{1},\mathbf{p}^n_{2},...,%
\mathbf{p}^n_{N}]$ specifying $n$-step transitional probabilities for random
walk on an undirected connected graph of $N$ nodes can always be represented
by the following spectral expansion 
\begin{equation}
\mathbf{P}^{n}=\sum_{k=1}^{N}\lambda _{k}^{n}\mathbf{v}^{k}\otimes \mathbf{a}%
^{k}\ \text{or}\ \mathbf{p}_{i}^{n}=\sum_{k=1}^{N}\lambda _{k}^{n}\mathbf{v}%
^{k}a_{i}^{k}  \label{PE2}
\end{equation}%
where 
\begin{equation}
\mathbf{Pv}^{k}=\lambda _{k}\mathbf{v}^{k},\ \ \ \mathbf{a}^{k}=\mathbf{D}%
^{-1}\mathbf{v}^{k}
\end{equation}%
\ so that the eigenvalues $\lambda _{k}$, the eigenvectors $\mathbf{v}^{k}$
and the spectral coefficients $\mathbf{a}^{k}$ are real. The eigenvectors
are orthonormal 
\begin{equation}
\left\langle \mathbf{v}^{k}\mathbf{,v}^{j}\right\rangle =I^{kj},\ \ \
I^{kj}=\left\{ 
\begin{tabular}{cc}
$1,$ & $k=j$ \\ 
$0,$ & $k\neq j$%
\end{tabular}%
\right. \   \label{orth}
\end{equation}%
in the sense of the inner product is defined by 
\begin{equation}
\left\langle \mathbf{v}^{k}\mathbf{,v}^{j}\right\rangle =\left( \mathbf{v}%
^{k}\right) ^{\func{T}}\mathbf{S\mathbf{v}}^{j}=\left\langle \mathbf{v}^{j},%
\mathbf{v}^{k}\right\rangle \mathbf{,\ \ \ S=S^{\func{T}}=\mathbf{D}^{-1}\ }
\label{innerp}
\end{equation}
\end{lemma}

According to definition (\ref{innerp}), the operator $\mathbf{P}$ is
self-adjoint, that is $\mathbf{P}^{\ast }\mathbf{=P}$ where the adjoint
operator $\mathbf{P}^{\ast }$ is defined in terms of the inner product $%
\left\langle \mathbf{x,Py}\right\rangle =\left\langle \mathbf{P}^{\ast }%
\mathbf{x,y}\right\rangle =\left\langle \mathbf{y,P}^{\ast }\mathbf{x}%
\right\rangle $ with arbitrary real $\mathbf{x}$ and $\mathbf{y}$. Indeed,
we may write%
\begin{eqnarray*}
\left\langle \mathbf{x,Py}\right\rangle &=&\mathbf{x}^{^{\func{T}}}\mathbf{%
\mathbf{D}^{-1}Py}=(1-\beta )\mathbf{x}^{^{\func{T}}}\mathbf{D}^{-1}\mathbf{%
AD}^{-1}\mathbf{y+}\beta \mathbf{x}^{^{\func{T}}}\mathbf{D}^{-1}\mathbf{y} \\
&\mathbf{=}&(1-\beta )\mathbf{y}^{^{\func{T}}}\mathbf{D}^{-1}\mathbf{A^{^{%
\func{T}}}D}^{-1}\mathbf{x+}\beta \mathbf{y}^{^{\func{T}}}\mathbf{D}^{-1}%
\mathbf{x=x}^{^{\func{T}}}\mathbf{\mathbf{D}^{-1}Py=}\left\langle \mathbf{%
y,Px}\right\rangle
\end{eqnarray*}%
for any $\mathbf{x}$ and $\mathbf{y}$ since $\mathbf{A}$ is symmetric ($%
\mathbf{A^{^{\func{T}}}=A}$) and $\mathbf{D}$ is diagonal. Self-adjoint
compact operators are subject to the Hilbert--Schmidt theorem (being defined
in a Euclidean space of finite dimension, the operator $\mathbf{P}$ is
always compact -- see \citet{KolmFomin}). Hence, according to this theorem,
the eigenvalues $\lambda _{k}$ are real ($\left\vert \lambda _{k}\right\vert
=\lambda _{k}$) and eigenvectors $\mathbf{v}^{k}$ can be chosen real and
orthonormal (in the sense of the inner product defined by (\ref{innerp}) but
not in the sense of the dot product used in (\ref{S1})). Finally we note
that operation $a^{k}=\left\langle \mathbf{v}^{k}\mathbf{,y}\right\rangle =(%
\mathbf{v}^{k})^{\func{T}}\mathbf{D}^{-1}\mathbf{y}$ determines coefficients
for spectral expansion $\mathbf{y=\Sigma }_{k}\mathbf{v}^{k}a^{k}$ of an
arbitrary\ vector $\mathbf{y}$ in the basis of $\mathbf{v}^{k}$. Hence $%
\mathbf{Py=\Sigma }_{k}\lambda _{k}\mathbf{v}^{k}a^{k}$. Since this basis is
complete, the transition operator $\mathbf{P}$ can be expressed in terms of
the outer product by 
\begin{equation}
\mathbf{P=}\sum_{k=1}^{N}\lambda _{k}(\mathbf{v}^{k}\otimes \mathbf{v}^{k})%
\mathbf{D}^{-1}=\sum_{k=1}^{N}\lambda _{k}\mathbf{v}^{k}\left( \mathbf{a}%
^{k}\right) ^{\func{T}}  \label{PE1}
\end{equation}%
Applying operator $\mathbf{P}$ $\ n$ times, we obtain (\ref{PE2}). Expansion
(\ref{PE2}) is unique as long as the eigenvalues are not repeated.

It is easy to see that matrices $\mathbf{T}$ and $P$ have the same set of
eigenvectors $\mathbf{Tv}^{k}=\mu _{k}\mathbf{v}^{k}$ and $\mathbf{Pv}%
^{k}=\lambda _{k}\mathbf{v}^{k}$ but shifted eigenvalues $\lambda
_{k}=(1-\beta )\mu _{k}+\beta $. Since the absolute values of eigenvalues of 
$\mathbf{T}$ are bounded by unity $\left\vert \mu _{k}\right\vert \leq 1$
(the matrix $\mathbf{T}$ must preserve the overall probability $\Sigma
_{i}p_{i}=1)$, the choice of $\beta =1/2$ enforces non-negativeness of the
eigenvalues $0\leq \lambda _{k}\leq 1$. This underpins predominately
monotonic convergence $\mathbf{p}_{i}^{n}\rightarrow \mathbf{p}^{\circ }$ as 
$n\rightarrow \infty $ (as it is shown above, eigenvalues $\lambda _{k}$\
must be real $\lambda _{k}=\left\vert \lambda _{k}\right\vert $). Since the
graph is presumed to be connected, there exists $n_{0}$ so that all
transitional probabilities are strictly positive $P_{ji}^{n}>0$ for any $%
n\geq n_{0}$. Hence, the transition matrix is subject to the conditions of
the Perron--Frobenius theorem \citep[see][]{Gantmakher} so that its largest
eigenvalue is positive, distinct and must be $\lambda _{1}=1$ to preserve
the overall probability. Hence, without loss of generality, we presume in
the rest of the paper that the eigenvalues are ordered 
\begin{equation}
1=\lambda _{1}>\lambda _{2}\geq ...\geq \lambda _{N}\geq 0  \label{lams}
\end{equation}%
Any quantities related to a selected group of the largest lambdas (e.g. $%
\lambda _{1},...,\lambda _{k},\ \ \ k<N$) are be referred to as "leading".

We can define our understanding of clusters in terms of the spectral
expansions specified by (\ref{PE2}). In its spectral form, the clustering
assumption becomes 
\begin{equation}
a_{i}^{k}=a_{\mathfrak{C}}^{k}\ \ \text{for}\ \ k\leq m(\mathfrak{C)}\ \text{%
and}\ i\in \mathfrak{C}  \label{ClustCond}
\end{equation}%
that is for any node $i$ that belongs to cluster $\mathfrak{C}$ there exist
such $m$ dependent on $\mathfrak{C}$ that for any $k\leq m$ the spectral
coefficient are independent of $i$. The coefficients $a_{\mathfrak{C}%
}^{1},...,a_{\mathfrak{C}}^{m}$ are thus spectral characteristics of the
cluster $\mathfrak{C}$ and would be different for a different cluster.
Hence, the powers of the transition operator can be represented by 
\begin{equation}
\mathbf{p}_{i}^{n}=\mathbf{p}_{\mathfrak{C}}^{n}+\mathbf{g}_{i}^{n}(%
\mathfrak{C}),\ \ i\in \mathfrak{C}\   \label{PP}
\end{equation}%
where%
\begin{equation}
\mathbf{p}_{\mathfrak{C}}^{n}=\sum_{k=1}^{m}\lambda _{k}^{n}\mathbf{v}^{k}a_{%
\mathfrak{C}}^{k},\ \ \mathbf{g}_{i}^{n}(\mathfrak{C})=\sum_{k=m+1}^{N}%
\lambda _{k}^{n}\mathbf{v}^{k}a_{i}^{k}\sim O\left( \lambda _{m+1}^{n}\right)
\end{equation}%
We note that, according to this definition, all nodes belong to the overall
graph cluster $\mathfrak{G}$ that has $m=1$ and involves all nodes of the
graph: $a_{i}^{1}=a_{\mathfrak{G}}^{1}$ for all $i=1,...,N$. Each cluster $%
\mathfrak{C}$ is associated with the characteristic time 
\begin{equation}
t_{\mathfrak{C}}=\frac{1}{\left\vert \ln \lambda _{m+1}\right\vert }
\end{equation}%
so that $\mathbf{g}_{i}^{n}(\mathfrak{C})$ is exponentially small for $t\gg
t_{\mathfrak{C}}$, no larger than $\sim \exp \left( -t/t_{\mathfrak{C}%
}\right) $. \ We denote $t=n$ to clearly outline the time dependence. A
cluster $\mathfrak{C}$ should appear well before the stationary
distributions are established to be distinguishable from the overall cluster 
$\mathfrak{G}$. Hence it is expected that $t_{\mathfrak{C}}\ll t_{\mathfrak{G%
}}=1/\left\vert \ln \lambda _{2}\right\vert $ for any cluster $\mathfrak{C}$
distinguishable from $\mathfrak{G}$. Here, $t_{\mathfrak{G}}$ is the
characteristic time\ of achieving steady-state distributions in the whole
network.

In practice, however, condition (\ref{ClustCond}) is not satisfied exactly
for any $m>1$. Only when $m=1,$ all of $a_{1}^{1},...,a_{N}^{1}$ are exactly
the same to specify the stationary solution. This implies that equation (\ref%
{PP}) needs to be corrected for deviations of $a_{i}^{k}$ from $a_{\mathfrak{%
C}}^{k},$ which nevertheless are expected to be small. To reflect this,
condition (\ref{ClustCond}) is replaced by 
\begin{equation}
\left\vert a_{i}^{k}-a_{\mathfrak{C}}^{k}\right\vert \sim \varepsilon \ll 1\ 
\text{for}\ k\leq m(\mathfrak{C)}\ \text{and}\ i\in \mathfrak{C}
\end{equation}%
We therefore obtain: 
\begin{equation}
\mathbf{p}_{i}^{n}=\mathbf{p}_{\mathfrak{C}}^{n}+\mathbf{g}_{i}^{n}(%
\mathfrak{C})+\mathbf{h}_{i}^{n}(\mathfrak{C}),\ \ i\in \mathfrak{C}\ 
\end{equation}%
where%
\begin{equation}
\mathbf{h}_{i}^{n}(\mathfrak{C})=\sum_{k=2}^{m}\lambda _{k}^{n}\mathbf{v}%
^{k}\left( a_{i}^{k}-a_{\mathfrak{C}}^{k}\right) \sim \varepsilon O\left(
\lambda _{2}^{n}\right) =\varepsilon O\left( \exp \left( -\frac{t}{t_{%
\mathfrak{G}}}\right) \right)
\end{equation}%
reflects spectral imperfections in representation of the clusters.

Our analysis becomes more transparent if the following interpretation of the
similarity criterion is used:%
\begin{equation}
R_{ij}^{n}=\frac{\left\langle \mathbf{p}_{i}^{n}\mathbf{,p}%
_{j}^{n}\right\rangle }{\left\Vert \mathbf{p}_{i}^{n}\right\Vert \left\Vert 
\mathbf{p}_{j}^{n}\right\Vert },\ \ \ R_{ij}^{n}\geq 1-\delta \ \
\Longrightarrow \ \ \mathbf{p}_{i}^{n}\cong \mathbf{p}_{j}^{n}  \label{Rdel}
\end{equation}%
where\ $\left\Vert \mathbf{x}\right\Vert \equiv \left\langle \mathbf{x,x}%
\right\rangle ^{-1/2},\ \ \left\langle \mathbf{x,y}\right\rangle =\mathbf{x}%
^{\func{T}}\mathbf{Sy}$ for any $\mathbf{x}$ and $\mathbf{y}$. If $\mathbf{p}%
_{i}^{n}=\mathbf{p}_{\mathfrak{C}}^{n}+$\textbf{$f$}$_{i}^{n}$ then we can
expand assuming that variations \textbf{$f$}$_{i}^{n}$ are small%
\begin{equation}
R_{ij}^{n}=1-\frac{1}{2}\frac{\left\langle \mathbf{f}_{i}^{n}-\mathbf{f}%
_{j}^{n},\mathbf{f}_{i}^{n}-\mathbf{f}_{j}^{n}\right\rangle }{\left\langle 
\mathbf{p}_{\mathfrak{C}}^{n}\mathbf{,p}_{\mathfrak{C}}^{n}\right\rangle }+%
\frac{1}{2}\frac{\left\langle \mathbf{p}_{\mathfrak{C}}^{n},\mathbf{f}%
_{i}^{n}-\mathbf{f}_{j}^{n}\right\rangle ^{2}}{\left\langle \mathbf{p}_{%
\mathfrak{C}}^{n}\mathbf{,p}_{\mathfrak{C}}^{n}\right\rangle ^{2}}+O\left(
\left\Vert \mathbf{f}_{i}^{n}-\mathbf{f}_{j}^{n}\right\Vert ^{3}\right)
\label{Rexp}
\end{equation}%
To avoid unnecessary complexities, we put $\mathbf{S=D}^{-1}$ ensuring
orthogonality of $\mathbf{g}_{i}^{n}(\mathfrak{C})$ and $\mathbf{h}_{i}^{n}(%
\mathfrak{C})$. That is $\left\langle \mathbf{g}_{i}^{n}\mathbf{,h}%
_{j}^{n}\right\rangle =0$ for any $i,j\in \mathfrak{C}$ and also $%
\left\langle \mathbf{g}_{i}^{n}\mathbf{,p}_{\mathfrak{C}}^{n}\right\rangle
=0 $ for any $i\in \mathfrak{C}$. Substitution of $\mathbf{f}_{i}^{n}=%
\mathbf{g}_{i}^{n}+\mathbf{h}_{i}^{n}$ into (\ref{Rexp}) yields 
\begin{eqnarray}
R_{ij}^{n} &=&1-\frac{1}{2}\frac{\left\Vert \mathbf{g}_{i}^{n}-\mathbf{g}%
_{j}^{n}\right\Vert ^{2}+\left\Vert \mathbf{h}_{i}^{n}-\mathbf{h}%
_{j}^{n}\right\Vert ^{2}}{\left\Vert \mathbf{p}_{\mathfrak{C}%
}^{n}\right\Vert ^{2}}+\frac{1}{2}\frac{\left\langle \mathbf{p}_{\mathfrak{C}%
}^{n},\mathbf{h}_{i}^{n}-\mathbf{h}_{j}^{n}\right\rangle ^{2}}{\left\Vert 
\mathbf{p}_{\mathfrak{C}}^{n}\right\Vert ^{4}}+...  \notag \\
&=&1-O\underset{\sim \lambda _{m+1}^{2n}}{\underbrace{\left( \exp \left( -2%
\frac{t}{t_{\mathfrak{C}}}\right) \right) }}-\varepsilon ^{2}O\underset{\sim
\lambda _{2}^{2n}}{\underbrace{\left( \exp \left( -2\frac{t}{t_{\mathfrak{G}}%
}\right) \right) }}
\end{eqnarray}%
for $i,j\in \mathfrak{C\ }$and $t\rightarrow \infty $. This representation
imposes restrictions on the choice of the threshold $\,\delta $ in (\ref{S1}%
). Indeed, on one hand we wish to set $\,\delta $ sufficiently small to
ensure accurate representation of the clusters. On the other hand, if $%
\lambda _{m+1}^{n}\sim \,\delta $ then $\lambda _{2}^{n}\gg \lambda
_{m+1}^{n}$ and, generally, $\lambda _{2}^{n}$ should be treated as being of
the order of unity since, as discussed previously, $t_{\mathfrak{G}}\gg t_{%
\mathfrak{C}}$. Hence, we should assume that the last term is $\sim
\varepsilon ^{2}$ at times $t\gtrsim t_{\mathfrak{C}}$ when cluster $%
\mathfrak{C}$ is being detected and select the threshold $\delta $\ within
the range 
\begin{equation}
1\gg \,\delta \gg \varepsilon ^{2}
\end{equation}%
to avoid interference of the spectral imperfections with the similarity
criterion given in (\ref{Rdel}).

The spectral understanding of clustering can be summarised in the following
proposition

\begin{proposition}
A cluster $\mathfrak{C}$ is a group of nodes that have the same or similar
values of $m$\ leading spectral coefficients, i.e. $a_{i}^{k}\approx
a_{j}^{k}$ where $i,j\in \mathfrak{C},$ $k=1,2,...,m$ and $\mathbf{a}^{k}=%
\mathbf{D}^{-1}\mathbf{v}^{k}.$ Different clusters have different spectral
coefficients and, generally, different values of $m$. Each cluster $%
\mathfrak{C}$ is associated with a certain characteristic diffusion time
given by $t_{\mathfrak{C}}=\left\vert \ln \lambda _{m+1}\right\vert ^{-1}.$
\end{proposition}

Finally, it is useful to stress the relation between the temporal and
spectral properties used to define clusters:

\begin{proposition}
If $m$\ leading spectral coefficients are the same $a_{i}^{k}=a_{j}^{k}$ for
two nodes $i$ and $j$ and for $k=1,...,m,$ then $R_{ij}^{n}=1-O\left(
\lambda _{m+1}^{2n}\right) $ as $n\rightarrow \infty $. If $%
R_{ij}^{n}=1-O\left( \omega ^{2n}\right) $ as $n\rightarrow \infty $, for
two nodes $i$ and $j,$ and for any fixed value $0<\omega <1,$ then the
leading spectral coefficients must be the same $a_{i}^{k}=a_{j}^{k}$ for any 
$k$ that $\lambda _{k}>\omega $.
\end{proposition}

The first part of the statement immediately follows from the spectral
expansion in Lemma \ref{lem1} and is obvious. The second part can be easily
proven by assuming that $a_{i}^{k}\neq a_{j}^{k}$ and finding that the
result $R_{ij}^{n}=1-O\left( \lambda _{k}^{2n}\right) $ contradicts the
condition $R_{ij}^{n}=1-O\left( \omega ^{2n}\right) $ in the statement when $%
\lambda _{k}>\omega $. There is another possibility of $a_{i}^{k}=ca_{j}^{k}$
with constant $c\neq 1$ and $k=1,...,m$ where $\lambda _{m+1}\leq \omega ,$
which is compliant with $R_{ij}^{n}=1-O\left( \omega ^{2n}\right) $. This
case, however, is inconsistent with the stationary distribution requiring $%
a_{i}^{1}=a_{j}^{1}$.

\section{Clustering of industrial network}

The procedure specified in Section \ref{SecRW} is used here with $\beta =1/2$
in (\ref{Pbet})\ and $\delta =10^{-2}$ in (\ref{SC}). As noted in Section %
\ref{SecRW}, the binary relation of cluster similarity is subjected to
transitive closure to avoid dependence of clusters on initial ordering of
the nodes. Unless otherwise stated, the nodes are preordered using the
Fiedler vector. The graph of industrial product connections was constructed
following the procedure suggested by \citet{Atlas2007}. The data are taken
from the United Nation Comtrade website using SITC2 classification at
4-digit level. This specifies $770$ categories of products traded by 153
countries. The graph with $N=770$ nodes is then constructed by Kruskal's
algorithm followed by adding another 1000 connections that have highest
proximities. The details can be found in thesis by \citet{Bouet2018}.\ All
these steps and parameters are consistent with those selected by %
\citet{Atlas2007} and \citet{Atlas2014}. The results presented below are
generally stable with respects to the selection of the parameters.

A similar group of nine primary clusters tends to appear irrespective of the
variations in the clustering procedure, although some details and the final
ordering may vary. The cluster structure is shown by clustering maps in
Figure \ref{fig1}. Primary clusters, which have the shortest characteristic
times, tend to merge into secondary clusters that have longer characteristic
times and, after several rounds of merges, form the overall cluster $%
\mathfrak{G}$ that covers all of the nodes. The two maps shown correspond to
different initial (and final) orderings, according to SITC classification on
the left and using the Fiedler vector on the right. Yet the clusters, which
are indicated by numbers and red lines, are exactly the same. Ordering of
elements on the maps does not change with time: each of\ the clustering maps
uses the corresponding final ordering produced by the clustering algorithm.
Figure \ref{fig1} also indicates the characteristic time scales of cluster
formation, growth and merging. It can be seen that, in general, cluster
compositions evolve with time and thus can be selected differently (in
Figure \ref{fig1}, out selection is indicated by vertical lines). We
nevertheless expect that a well-defined cluster exists in a fixed or
slightly changing boundaries over some range of characteristic times.

The determined primary clusters are briefly described in Table 1. Many of
the clusters (at least 3 "mechanical" clusters) are associated with
machinery. Two of these clusters (Machinery-2 and Machinery-3) merge early
to form the secondary Machinery cluster. Machinery-3 is more related to
construction than Machinery-1. The third machinery cluster (Machinery-1)
seems to be more closely associated with general chemical industry
(Chemicals-1). It is interesting that the second chemical cluster
(Chemicals-2), which is more specialised and less related to Machinery than
the first chemical cluster, takes longer to appear. Food industry (Food) and
building materials industry (Construction) appear to be closely related
through. Garments form a very distinctive and quite independent cluster
involving many related products. Mining and resources hardly form an
independent cluster and are distributed between other clusters.\bigskip

The clustering algorithm used in the present work forms the same clusters
irrespective of the initial ordering. These clusters are formed not due to
logical ordering of products by SITC classification (and, as discussed
below, are initially reordered by the Fiedler vector) but through
connections between industry products. Figure \ref{fig3} displays the same
adjacency matrix $\mathbf{A}$ with different ordering of the nodes: a)
according to SITC classification, b) using the Fiedler vector, c) using
clustering algorithm on original SITC ordering and d) using clustering
algorithm after pre-ordering by the Fiedler vector. The clustering algorithm
of the present work identifies some "fast" clusters that are missing by the
sorting using the Fiedler vector since the latter pertains to the divisions
associated with the slowest relaxation to the steady-state distribution.

Some of the products may be found in clusters that are seemingly not related
or related by practice rather than by a common origin or a conceptual link.\
For example many packaging products can be found not in Chemicals but in the
Food cluster, where these products are predominately used. Various resources
and materials can often be found in industrial clusters that use them.
Figure \ref{fig4} compares SITC ordering of the products with ordering
achieved by the clustering algorithm. The colour code corresponds to 1-digit
SITC codes. The nine primary clusters are indicated by the vertical bars.
The correspondence between the identified clusters and 1-digit codes is
apparent. \ While good ordering is important for visualisation, Figures \ref%
{fig4} and \ref{fig5} use initial odering ordering by the Fiedler vector and
not by the SITC codes. This is to make sure that the clustering algorithm is
not aware about the relationships between products, which is indicated by
the standard ordering of the SITC codes. Effectively, the clustering
algorithm cannot benefit from knowing the SITC codes and has to introduce
its own classification of the products from properties of the industrial
network. The classification by the algorithm does not match the SITC codes
exactly but appears to correlate with them. This correlation can be improved
by using 2-digit SITC codes and mapping them onto 8 categories that do not
coincide with the 1-digit SITC codes. The SITC codes reflect formal
classification that is traditionally used in trade statistics but does not
necessarily correspond to the role that these products play in the real
world. Figure \ref{fig5} demonstrates that this new refined coding improves
characterisation of the clusters. As shown in Figure \ref{fig6}, the refined
coding is achieved in two stages: first, mapping is performed on the basis
of the 1-digit codes as indicated by the thick arrows and, second, some of
the 2-digit categories of products are remapped according to our
understanding of their roles as shown by the thin arrows.

Finally, Figure \ref{fig7} demonstrates spectral representation for selected
clusters. Each subfigure characterises a single cluster and plots 20 lines
with each line corresponding to a distinct node from the cluster. The lines
are not similar in the top subfigure \ --- this subfigure does not
correspond to any cluster and is shown for comparison. The lines of the
bottom subfigures match much better than those of the top subfigures. The
effective value of $m$ increases and the corresponding characteristic time $%
t_{\mathfrak{C}}$ decreases from top to bottom. Hence, lower subfigures
correspond to faster forming clusters with shorter characteristic times.
This is consistent with cluster properties shown in Figure \ref{fig1}. We
also note that the first 5-10 modes are similar for the clusters Chemicals-2
and Electrical but different from the corresponding modes determined for the
cluster Garment. This is expected since clusters Chemicals-2 and Electrical
merge to form a secondary cluster.

\section{Discussion of the findings}

Absence of an autonomous cluster representing resources and mining seems
puzzling. Some of the simulations produced a cluster resembling mining, but
its detection was not reliable and such sporadically appearing clusters are
not shown in the present work (detection of clusters that have long
characteristic diffusion times cannot be reliable). Two factors can be
responsible for this behaviour. The first is strong linking of resources
with the industries that use these resources; the second is that mining and
resources are subject to geological constraints and thus may be strongly
affected by non-economic factors, which are not considered in the theory due
to \citet{Atlas2007}.

The time dependency of the clustering algorithm is essential for our
analysis. This time, however, reflects only the rate of diffusion of
innovation and should not be confused with the real physical time. While the
specific shapes of the detected clusters may be to some extent unexpected,
the main findings of this work tend to agree with what is generally known
about industrial development \citep{Atlas2007}. This confirms the approach
to clustering implemented in the present work. Industrial evolution forms
clusters of industries that tend to grow jointly in dynamic connection with
each other. Developing industries associated with clusters that have more
cohesive structure and smaller characteristic diffusion times is easier
since these clusters require less diverse resources and connections. The
garment industries are the fastest to develop, followed by agroindustry,
construction and machinery. Industries producing complex chemicals and
electronics tend to develop slower. The chemical industry has two quite
independent clusters: traditional chemistry (Chemicals-1), which develops
faster, and advanced chemistry (Chemicals-2), whose development takes a
longer time. This is consistent with commonly known economic trends: many
developing countries start from garment and agricultural industries and
develop heavier industries at later stages. In some newly-developed
economies, electronics may be one of the early additions to the industrial
mix but this seems to be the effect of government intervention.

The present consideration is based on modern trade data, which cannot
accurately reflect links between industries that existed many decades ago.
Yet, there is something common present in the two and half centuries-long
industrial development of the world. The five Kondratiev waves of economic
growth, which were recognised by Kondratiev, conceptualised by Schumpeter
and documented by Freeman, were re-evaluated by Perez as surges of
technological innovation. These surges are initiated by technological
breakthroughs and drive subsequent waves of economic growth. These
technological surges, which are schematically presented in Figure \ref{fig8}%
, are roughly consistent with the diffusional speeds of cluster formation
obtained in the present work. The British revolution in textile production
is followed by surges of innovation in machinery and construction, and only
then by radical changes in chemistry and electronics. The data
characterising the present state of industrial production cannot possibly
reproduce the history of industrial development, but it seems that these
data reflect some technological fundamentals that tend to persist over time.
Modern industry is still not homogeneous and forms clusters of related
products. Therefore, we can and, in fact, should expect a forthcoming sixth
surge in industrial development, which seems to be emerging and is likely to
be related to communications, transport, automation, AI, commercial space
exploration, advanced materials and medicine, new sources of energy as well
as artificial intelligence, advanced knowledge and progressive
education.\bigskip

\section{Conclusions}

This work suggests an alternative implementation of the ideas, which were
introduced by \citet{Lamb2009}, to examine a hierarchy of network
communities by a random walk. This implementation involves a resorting
algorithm combined with identification of clusters on the basis of
similarity of the transitional properties of random walks originated at the
nodes forming a cluster. Different diffusion times correspond to clusters of
different levels. In spectral representation, the same property is reflected
in similarity of the leading spectral coefficients. While this work
endeavours to accurately define clusters, we must note that, by their
nature, clustering properties are not exact and always leave some freedom in
defining what similarity means in exact terms. Our interpretation of
clusters reflects physical understanding of industrial links introduced and
evaluated by \citet{Atlas2007} and \citet{Atlas2014}.

The clusters considered here involve a characteristic time scale as one of
the properties of every cluster. The presence of a time scale in the
analysis is consistent with the understanding of diffusion of industrial
technology introduced by \citet{Atlas2007}. This understanding links the
rate of innovation to the rate of diffusion of randomly walking particles on
a graph representing connections between industrial products. The present
analysis identifies around a dozen clusters of industrial products and is
consistent with the notion that modern industry forms well-connected
technological clusters. Conceptual agreement with known historical trends in
evolution of industries supports the thesis about the leading role of
technological progress in long-term economic changes.

\newpage

\bigskip


  \bibliographystyle{imamat}
\bibliography{clust}

\newpage
\begin{tabular}{|l|p{2cm}|p{11cm}|}
\hline
\textbf{No} & \textbf{Cluster \newline
Label} & \multicolumn{1}{|c|}{\textbf{Main products associated with the
cluster}} \\ \hline
\textbf{1} & Chemicals-1 & Common chemicals (organic and non-organic), some
processed materials and related machinery \\ 
\textbf{2} & Machinery-1 & Mostly machinery, often specialised, some
products related to chemicals, paper and food processing \\ 
\textbf{3} & Machinery-2 & General machinery and equipment (a few are
related to transport, energy and military), measuring and controlling
devices, some supplies and materials. \\ 
\textbf{4} & Machinery-3 & Machinery and related materials, a few are
related to construction and agriculture \\ 
\textbf{5} & Construction & Construction materials and equipment, some
agricultural and household products, furniture \\ 
\textbf{6} & Food & Foods, drinks, tobacco, feeds and other agricultural
products, packaging \\ 
\textbf{7} & Chemicals-2 & Mixture of chemicals, gases and materials
(including industrial and for a special use, such as art, image \& photo
products), chemical and nuclear reactors \\ 
\textbf{8} & Electrical & Electrical machinery and equipment, electronics,
telecommunications, digital processors, circuits and controls, photo,
optical and other related equipment, some related materials \\ 
\textbf{9} & Garment & Closing, footwear, personal items, some fabrics and
related processed materials \\ \hline
\end{tabular}

\begin{center}
Table 1. Brief specification of the primary industrial clusters.
\end{center}

\bigskip

\begin{center}

\begin{figure}[p]
\begin{center}
\includegraphics[width=17cm,page=1, angle=180,origin=c,trim=1cm 3cm 1cm 5cm, clip ]{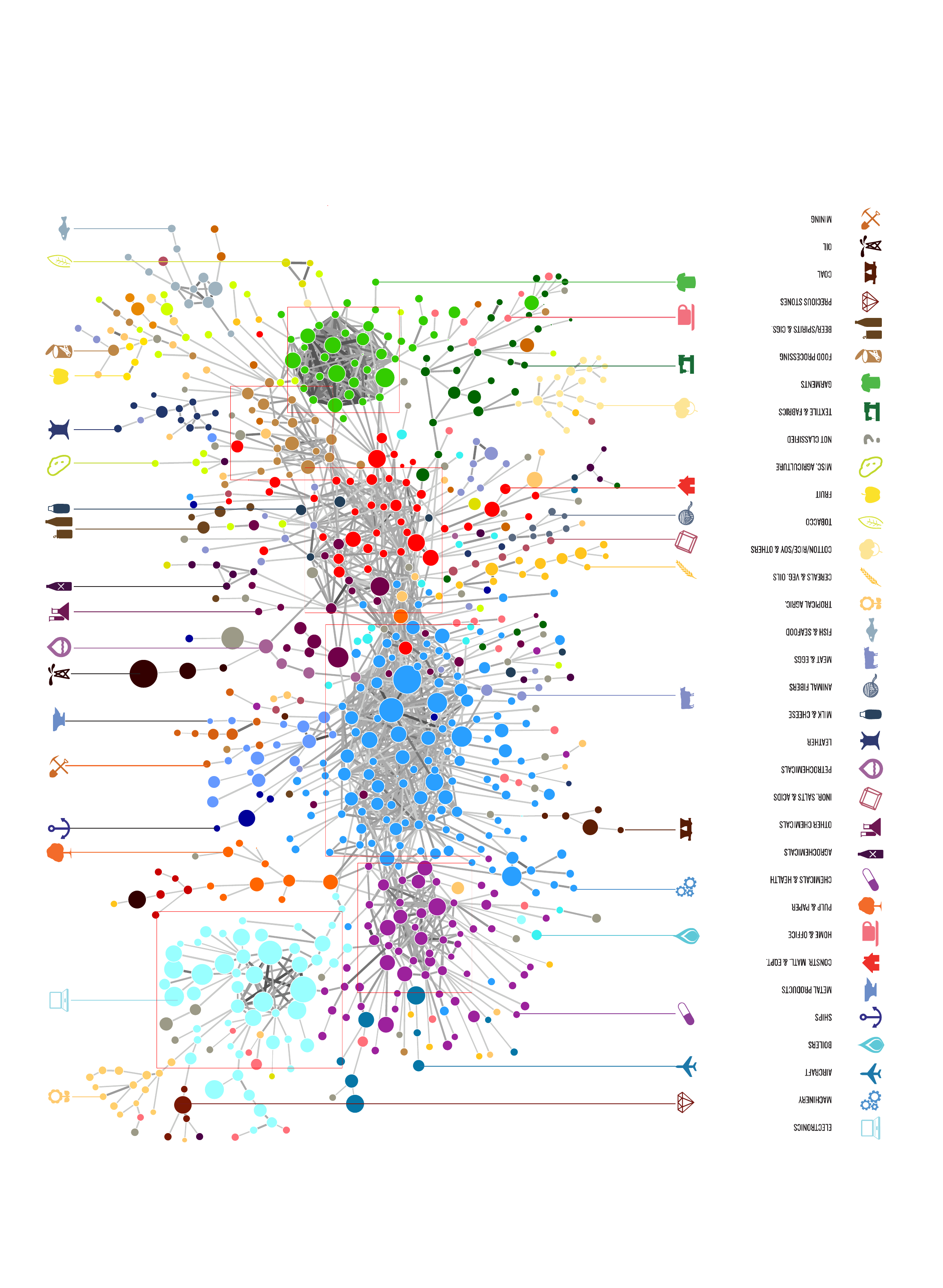}
\end{center}
\caption{Connections between industrial producs due to \citet{Atlas2014} (thanks: Cesar A. Hidalgo). The rectangles outline 
the core segments of the visually apparent clusters.}
\label{figa}
\end{figure}

\bigskip

\begin{figure}[h!]
\begin{center}
\includegraphics[width=\textwidth,page=1,trim=0cm 3.4cm 0cm 4.2cm, clip ]{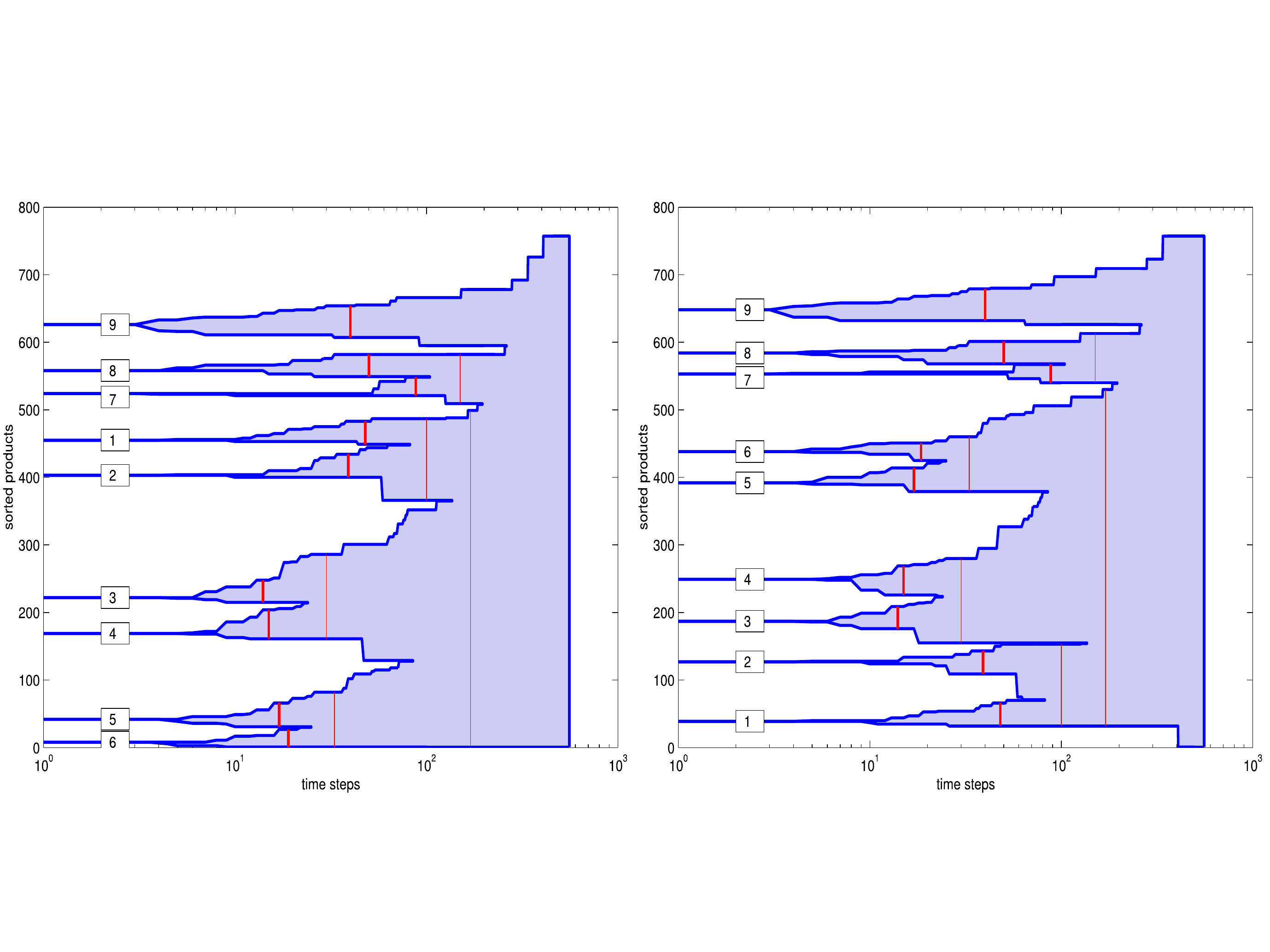}
\end{center}
\caption{Clustering map: extent of clusters of industrial products versus time step. The vertical lines show locations 
where the identified clusters are sampled (to be shown in the other figures; thick lines correspond to primary clusters). 
The initial ordering is according to SITC classification (left) or using the Fiedler vector (right).}
\label{fig1}
\end{figure}

\begin{figure}[h!]
\begin{center}
\includegraphics[width=18cm,page=1,trim=1cm 0cm 2cm 13cm, clip ]{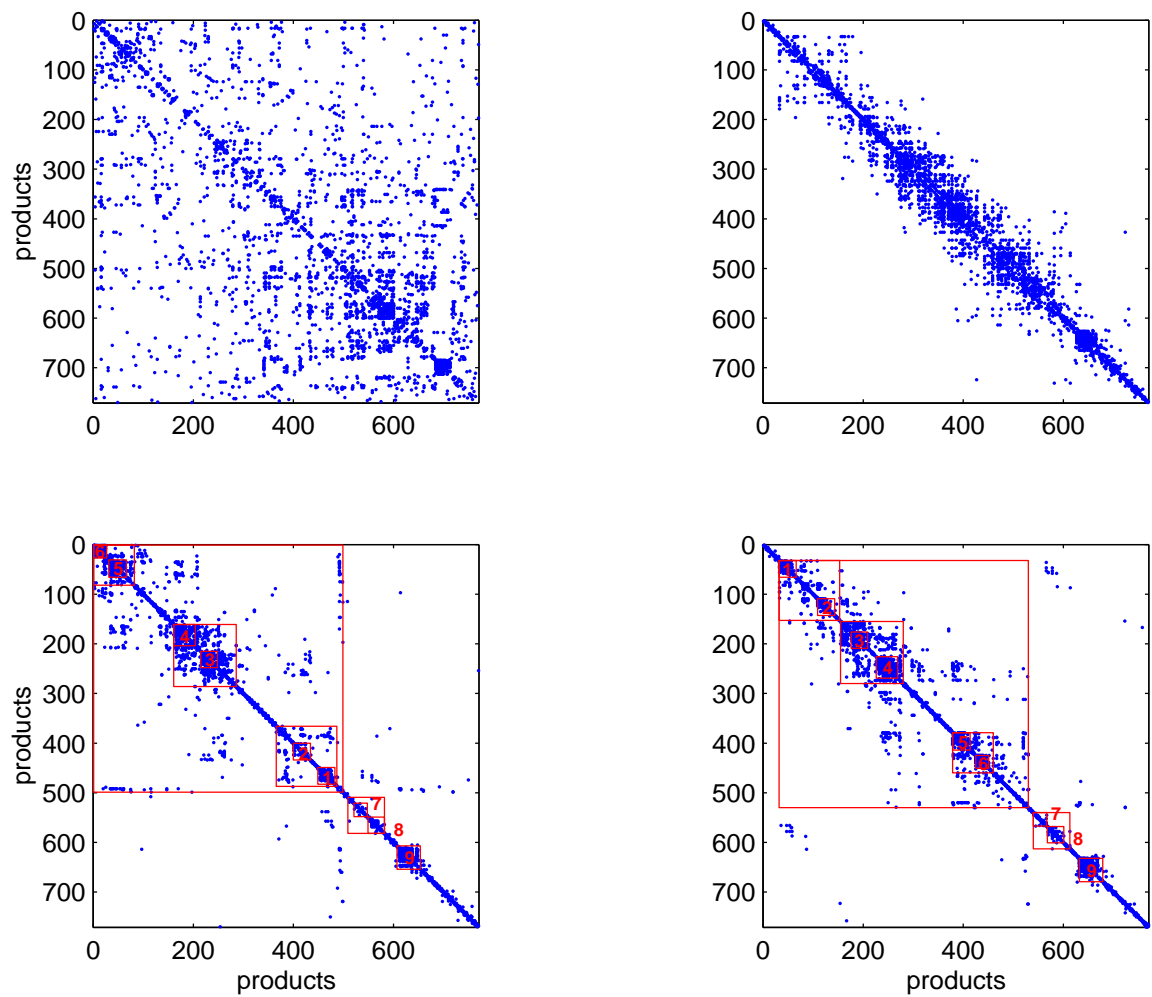}
\end{center}
\caption{Adjacency matrix shown with different sorting. Top row: no clustering, botom row: sorting by clustering algorithm. 
  Left column: (pre-)sorting according to SITC codes, right column:  (pre-)sorting by the Fiedler vector. The clusters identified in Figure \ref{fig2} are indicated by red squares.
 The numbers of the nine primary clusters are also shown. These clusters are the same in the two bottom figures and selected as 
shown in Figure 2.} 
\label{fig3}
\end{figure}

\vfill
\hspace{0pt}

\bigskip
\pagebreak
\vfill

\begin{figure}[h!]
\begin{center}
\includegraphics[width=12cm,page=1,trim=1cm 0cm 2cm 13cm, clip ]{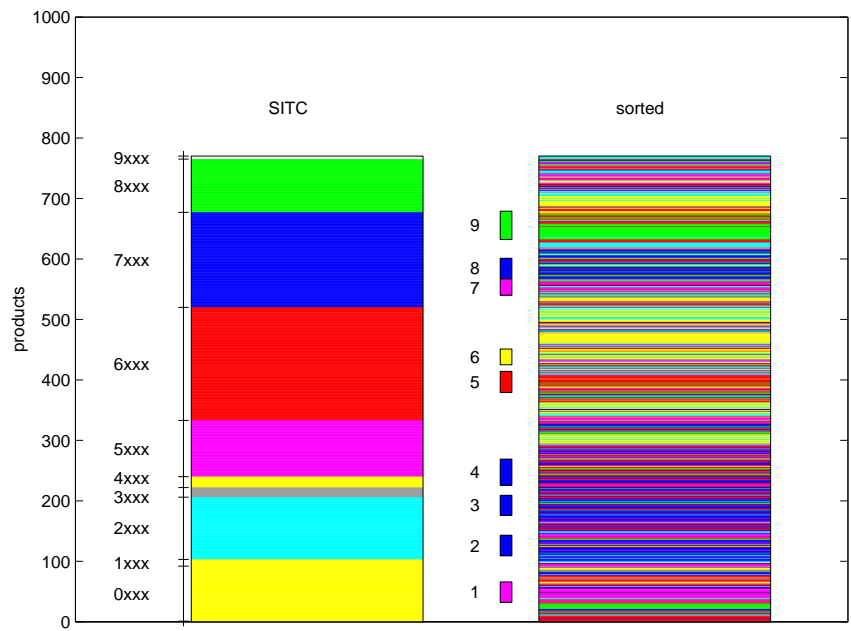}
\end{center}
\caption{Products ordered according to SITC codes (left) and ordered by the clustering algorithm (right). The vertical bars indicate the location of the 9 primary clusters. 
The colour code matches single-digit SITC. } 
\label{fig4}
\end{figure}

\begin{figure}[h!]
\begin{center}
\includegraphics[width=12cm,page=1,trim=1cm 0cm 2cm 13cm, clip ]{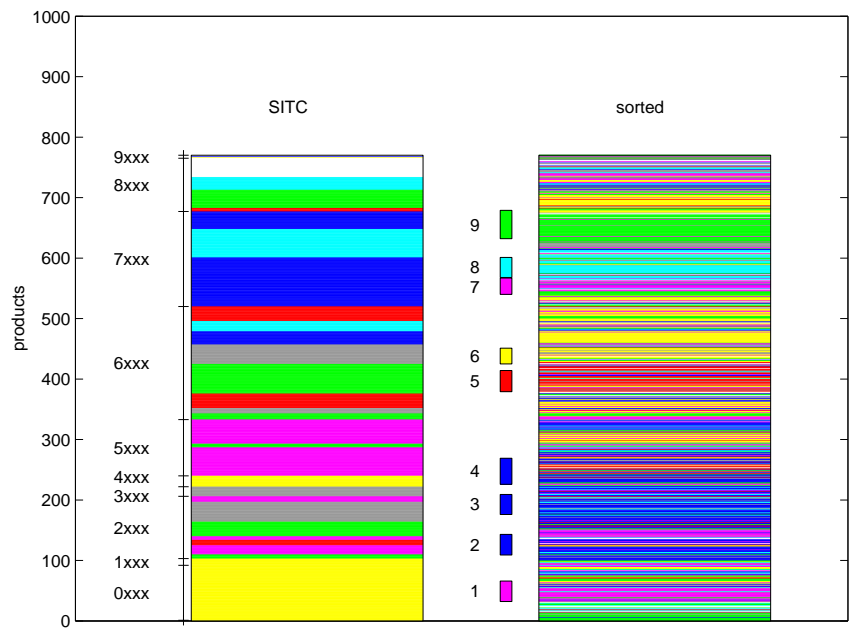}
\end{center}
\caption{The same as in the previous figure but with an alternative, refined colour coding.} 
\label{fig5}
\end{figure}

\begin{figure}[h!]
\begin{center}
\includegraphics[width=16cm,page=2,trim=0cm 0cm 0cm 0cm, clip ]{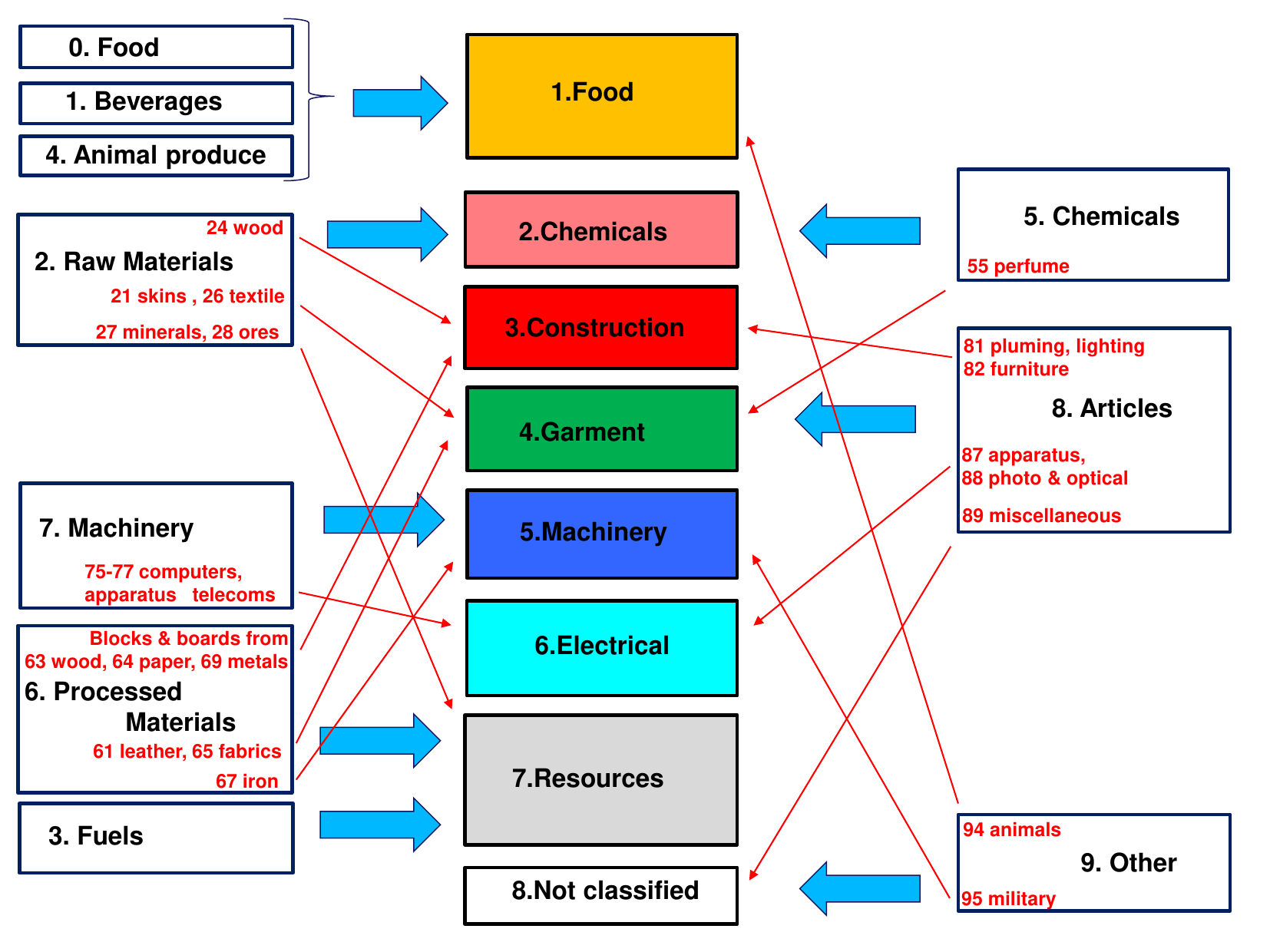}
\end{center}
\caption{Maping of SITC codes to cluster-related classification introducing the refined color coding, which is used in the previous figure.}
\label{fig6}
\end{figure}

\begin{figure}[h!]
\begin{center}
\includegraphics[width=16cm,page=1,trim=1cm 0cm 2cm 13cm, clip ]{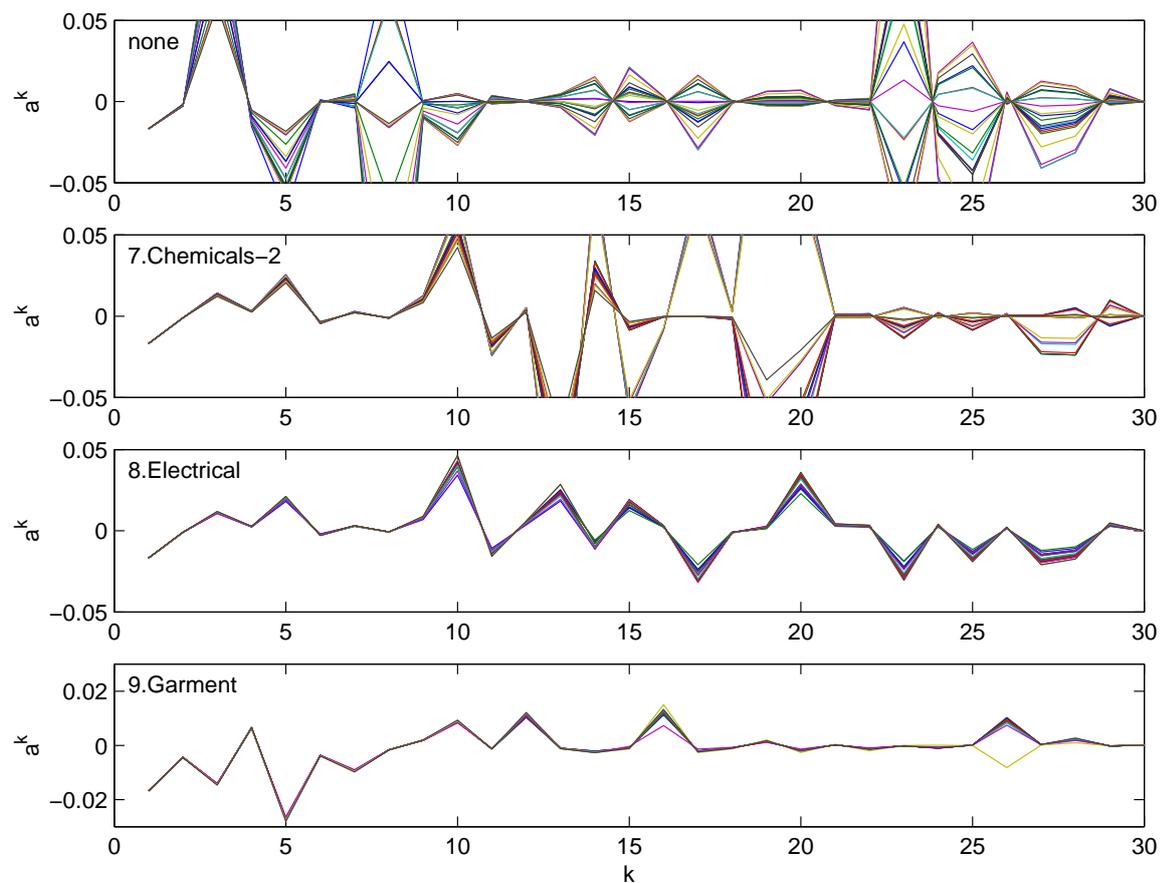}
\end{center}
\caption{Spectral representation for selected clusters. Each subfigure shows twenty lines corresponding to twenty nodes $i$ selected from a particular cluster plotted versus 
$k=1,2,...,30$. The top subfigure is shown for nodes $i=11,12,...,30$ (numbered after sorting), which do not correspond to any identified cluster. This subfigure is given for comparison.}
\label{fig7}
\end{figure}

\begin{figure}[h!]
\begin{center}
\includegraphics[width=16cm,page=3,trim=0cm 0cm 0cm 7cm, clip ]{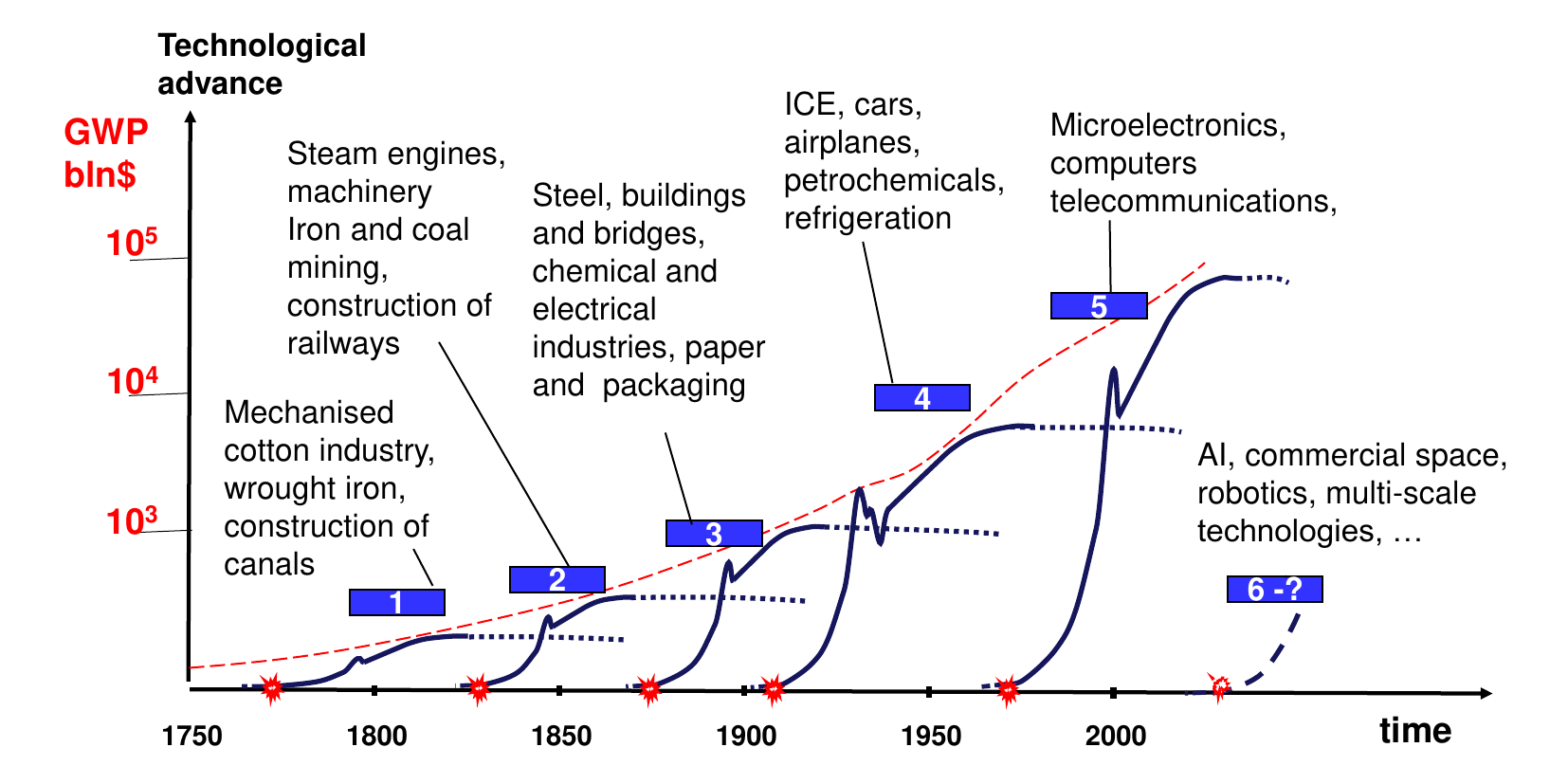}
\end{center}
\caption{Schematic of five technological surges due to \citet{Perez2006}. 
The figure also shows possible sixth surge and the estimate of Gross World Product using logarithmic scale.}
\label{fig8}
\end{figure}

\end{center}

\pagebreak

\end{document}